\documentclass[11pt,showpacs,nofootinbib]{revtex4}

\usepackage{amsfonts}
\usepackage{amsbsy} 
\usepackage{epsfig}
\usepackage{latexsym}
\input amssym.def
\input amssym.tex

\def\half{{1\over 2}}
\def\ben{\begin{equation}}
\def\een{\end{equation}}
\def\bena{\begin{eqnarray}}
\def\nn{\nonumber}
\def\eena{\end{eqnarray}}

\renewcommand{\theequation}{\arabic{section}.\arabic{equation}}
\advance \topmargin by -\headheight
\advance \topmargin by -\headsep
\evensidemargin \oddsidemargin
\advance\hoffset by -3mm  
\advance\voffset by  8mm  

\def\dS{\mathcal{D}}
\def\dSG{SO(d,1)}

\def\P{\mathcal{P}}
\def\M{\mathcal{M}}
\def\bR{\Bbb R}

\def\b1{e^0}

\begin{document}
\title{Deformed General Relativity and Torsion}

\author{Gary W. Gibbons\footnote{\tt gwg1@damtp.cam.ac.uk}, Steffen Gielen\footnote{\tt sg452@damtp.cam.ac.uk}}
\affiliation{D.A.M.T.P., Cambridge University, Wilberforce Road, Cambridge CB3 0WA, U.K.} 

\begin{abstract}
We argue that the natural framework for embedding the ideas of deformed, or doubly, special relativity (DSR) into a curved spacetime is a generalisation of Einstein-Cartan theory, considered by Stelle and West. Instead of interpreting the noncommuting ``spacetime coordinates" of the Snyder algebra as endowing spacetime with a fundamentally noncommutative structure, we are led to consider a connection with torsion in this framework. This may lead to the usual ambiguities in minimal coupling. We note that observable violations of charge conservation induced by torsion should happen on a time scale of $10^3$ s, which seems to rule out these modifications as a serious theory. Our considerations show, however, that the noncommutativity of translations in the Snyder algebra need not correspond to noncommutative spacetime in the usual sense.
\end{abstract}


\maketitle

\section{Introduction}
\label{intro}

It is commonly assumed that quantum gravity sets a fundamental length scale, the Planck scale \cite{planck}, which can not be resolved by any physical experiment. Different approaches to quantum gravity, such as string theory or loop quantum gravity, incorporate such a scale. This leads to the idea that some kind of ``space discreteness" should be apparent even in a low-energy ``effective" theory.

The idea of putting quantum mechanics on a discrete lattice\footnote{with spacing equal to the Compton wavelength of the proton, $l_c\approx 1.3$ fm} seems to have been first considered by Heisenberg in the spring of 1930 \cite{heisenberg}, in an attempt to remove the divergence in the electron self-energy. Because the absence of continuous spacetime symmetries leads to violations of energy and momentum conservation, this approach was not pursued further, but later in the same year he considered modifying the commutation relations involving position operators instead \cite{heisenberg}. 

A fundamental length scale is absent in special relativity, where two observers will in general not agree on lengths or energies they measure. Hence the usual ideas of Lorentz and Poincar\'e invariance must be modified in some way. Snyder observed \cite{snyder} that this could be done by deforming the Poincar\'e algebra into the de Sitter algebra, i.e. considering the isometry group of a (momentum) space of constant curvature. From an algebraic viewpoint, if one maintains the structure of a Lie algebra and considers deformations of the Poincar\'e algebra, the de Sitter algebra is the unique way of implementing a modified kinematic framework \cite{bacry1}.

A $d$-dimensional de Sitter momentum space with curvature radius $\kappa$ is defined as the submanifold of a $(d+1)$-dimensional flat space with metric signature ($d,1$) by
\ben
(P^1)^2 + (P^2)^2 + \ldots + (P^{d-1})^2 - (P^d)^2 + (P^{d+1})^2 = \kappa^2\,,
\label{ads}
\een
where $\kappa$ has dimensions of mass. Its isometry group is generated by the algebra
\bena
[M_{ab},M_{cd}] =\eta_{ac}M_{bd}+\eta_{bd}M_{ac}-\eta_{bc}M_{ad}-\eta_{ad}M_{bc}\,, \nn
\\ \left[ X_{a},M_{bc} \right]=\eta_{ac}X_b-\eta_{ab}X_c\,, \quad [X_a,X_b]=\frac{1}{\kappa^2}M_{ab}\,.
\label{algebra}
\eena
Here $M_{ab}$ correspond to a Lorentz subalgebra of the de Sitter algebra, while $X_a\equiv\frac{1}{\kappa}M_{d+1,a}$ are interpreted as (noncommuting) translations. These translations are then interpreted as corresponding to coordinates on spacetime; Snyder thought of operators acting on a Hilbert space. Since the operators $X_1,X_2$ and $X_3$ correspond to rotations in the $(d+1)$-dimensional space, their spectrum is discrete. In this way, one obtains ``quantised spacetime", while maintaining Lorentz covariance.

One can give explicit expressions for the algebra elements by choosing coordinates on de Sitter space (\ref{ads}). The choice made by Snyder is taking Beltrami coordinates
\ben
p^1=\kappa\frac{P^1}{P^{d+1}}\,,\;p^2=\kappa\frac{P^2}{P^{d+1}}\,,\ldots\,,\;p^d=\kappa\frac{P^d}{P^{d+1}}\,,
\een
whence one has $(P^{d+1})^2=\kappa^4/(\kappa^2+\eta_{ab}p^a p^b)$ to satisfy (\ref{ads}), and $\eta_{ab}p^a p^b \ge -\kappa^2$, corresponding to an apparent maximal mass if $p^a$ were interpreted as Cartesian coordinates on a Minkowski momentum space. (Up to this point one could in principle have chosen anti-de Sitter instead of de Sitter space. Then this inequality becomes $\eta_{ab}p^a p^b \le \kappa^2$, which perhaps seems less motivated physically.) A necessary sign choice means that these coordinates only cover half of de Sitter space. In these coordinates, the translation generators
\ben
X_a=\frac{1}{\kappa}\left(P^{d+1}\frac{\partial}{\partial P^a}-P_a\frac{\partial}{\partial P^{d+1}}\right)=\frac{\partial}{\partial p^a}+\frac{1}{\kappa^2}p_a p^b\frac{\partial}{\partial p^b}
\een
generate ``displacements" in de Sitter space. (In this notation, indices are raised and lowered with $\eta_{ab}$, the $d$-dimensional Minkowski metric, so that $p_a=\eta_{ab}p^b$.)

The motivation behind these ideas was to cure the infinities of quantum field theory, which evidently arise from allowing arbitrary high momenta (or short distances). In a somewhat similar spirit, Gol'fand suggested \cite{golfand} to define quantum field theory on a momentum space of constant curvature, using Beltrami coordinates as momentum variables. This makes the volume of the corresponding Riemannian space finite and so presumably leads to convergent loop integrals in the Euclideanised theory. The consequences for standard quantum field theory were further explored in \cite{kada,golfand2}.

 Gol'fand only assumed that $\kappa\gg m$ for all elementary particles; thinking of quantum gravity, one would perhaps identify $\kappa$ with the Planck scale, whereas the original authors seem to have thought of the Fermi scale.

 The induced metric on de Sitter space in terms of the coordinates $p^a$ is
\ben
g_{nr}=\frac{\kappa^2}{\kappa^2+p\cdot p}\left(\eta_{nr}-\frac{p_n p_r}{\kappa^2+p\cdot p}\right)\,,
\label{metric}
\een
where $p\cdot p\equiv \eta_{cd}p^c p^d$. The metric (\ref{metric}) becomes singular when $p \cdot p\rightarrow -\kappa^2$, and negative definite when extended to what Gol'fand calls the exterior region $p \cdot p < -\kappa^2$. In four dimensions, 
\ben
\det g = -\kappa^{10}(\kappa^2+p\cdot p)^{-5}\,,
\een
and the volume element is $d^4 p\,\kappa^5(\kappa^2+p\cdot p)^{-5/2}$.

 In Gol'fand's approach (assuming $d=4$ of course), the standard Feynman rules were modified by replacing the addition of momenta $p$ and $k$ at a vertex by
\ben
(p(+)k)^a=\frac{\kappa}{\kappa^2 - p\cdot k}\left(p^a\sqrt{\kappa^2+k\cdot k}+k^a\left(\kappa-\frac{p\cdot k}{\kappa+\sqrt{\kappa^2+k\cdot k}}\right)\right)\,,
\een
which corresponds to a translation by $k$ of the vector $p$. (Again $p\cdot k \equiv \eta_{ab}p^a k^b$, etc.) It was also noted that spinors now transform under ``displacements" as well, which is made more explicit in \cite{kada} and \cite{golfand2}. As is well known, five-dimensional Dirac spinors still have four components and the matrix $\gamma^5$ appears in the Dirac Lagrangian, hence there is no chirality. This alone seems to imply that the original Gol'fand proposal cannot be used for an appropriate model of the known particles.

Gol'fand's approach is very different from more recent approaches to quantum field theory on noncommutative spaces (see e.g. \cite{nekrasov}) in that the field theory is defined on a momentum space which is curved, but neither position nor momentum space are noncommutative in the usual sense. 

In this paper, we attempt to embed the old idea of a curved momentum space into general relativity by describing a geometric framework in which an internal de Sitter space is associated to a curved spacetime. This internal space replaces the usual (co-)tangent space in general relativity. We will make use of the interpretation of Einstein-Cartan theory given by Stelle and West \cite{stellewest}. Since we are staying within conventional differential geometry, this formalism provides an alternative to the usual interpretation of the Snyder algebra as describing a noncommutative spacetime.

The paper is organised as follows: We give a brief introduction into the ideas of {\it deformed (doubly) special relativity} (DSR) most relevant to the following discussion in section \ref{dsr}. In section \ref{gauge} we outline how Einstein-Cartan theory can be formulated as a gauge theory of gravity with the de Sitter group $\dSG$ as gauge group; this theory includes a gauge field that plays a crucial role in what follows. In this section we essentially rederive the results of Stelle and West, using a different set of coordinates which we find more closely related to the DSR literature. Since we claim that this geometric framework can be used to generalise the ideas of DSR, we show in section \ref{synth} how, if spacetime is taken to be Minkowski space, the simplest non-trivial choice of zero section leads to a connection with torsion, providing a geometric interpretation for the noncommuting ``coordinates" appearing in the Snyder algebra. We close with a discussion of our results and their possible physical implications, which show that the theory, at least in its given form, is not physically viable. We conclude that there may be different physical interpretations of algebraic commutation relations such as those used in DSR.

Since the two most obvious extensions of general relativity are admitting either connections with torsion or non-metric connections, we briefly discuss the theory of a torsion-free non-metric connection, known as symmetric affine theory, in an appendix. It does not fit as well into a description by Cartan geometry as the case highlighted in this paper. A more mathematical account of Cartan geometry is given in a second appendix.

We use units in which $\hbar=c=1$, such that momenta have the dimension of inverse length. Lower-case Latin indices such as $a,b,c$ denote either Lorentz indices or label coordinates, as will hopefully be clear from the context.

\setcounter{equation}{0}
\section{Deformed Special Relativity}
\label{dsr}

The idea that the classical picture of Minkowski spacetime should be modified at small length scales or high energies was re-investigated in more recent times, motivated by the apparent existence of particles in ultra high energy cosmic rays whose energies could not be explained within special relativity \cite{experiment}. The proposed framework of deformed special relativity (DSR) \cite{amelino} modifies the Poincar\'e algebra, introducing an energy scale $\kappa$ into the theory, in addition to the speed of light $c$. This leads to a quantum ($\kappa$-)deformation of the Poincar\'e algebra \cite{majid}, with the parameter $\kappa$ associated with the newly introduced scale.

It was soon realised \cite{kowalski} that this deformed algebra is the algebra of the isometry group of de Sitter space, and that the symmetries of DSR could hence be obtained by identifying momentum space with de Sitter space, identifying $X_a$ as the generators of translations on this space. The constructions of DSR thus appear to be a resurrection of Snyder's and Gol'fand's ideas. We take this observation as the defining property of DSR, and will seek to describe a framework in which momentum space, or rather the (co-)tangent space in general relativity, is replaced by an ``internal" de Sitter space. We will see that this can best be done using Cartan geometry.

When discussing DSR as a modification of special relativity, we take the view that special relativity is defined as a kinematic framework with preferred inertial systems, related to one another by (proper) Lorentz transformations. That is, one has a flat spacetime on which there exist certain preferred coordinate systems, those in which the metric is diagonal with entries $\pm 1$. From this point of view, the choice of coordinates on the internal de Sitter space plays quite an important role if one is looking for a ``deformation" of special relativity including an energy scale $\kappa$. Such a deformation can only arise if the chosen coordinate system reduces to Cartesian coordinates on Minkowski space as $\kappa\rightarrow\infty$. The choice of coordinates is obviously not unique.

The generators of the algebra will take different explicit forms when different coordinate systems (on four-dimensional de Sitter space) are chosen. In \cite{kowalski} ``natural coordinates" are defined by, in the notation of section \ref{intro}, \footnote{Capital Latin indices such as $I$ and $J$ used in this section only run over spatial coordinates (from 1 to 3).}
\ben
g=\exp\left[p^I (M_{I4}+X_I)\right] \exp\left[p^4 X_4\right]\mathcal{O}\,,
\een
where $\mathcal{O}=(0,0,0,0,\kappa)$ is taken to be the origin of de Sitter space in five-dimensional Minkowski space, and $M_{I5}$ and $M_{45}$ correspond to translations in space and time. The coordinates one obtains are related to the five-dimensional coordinates by
\ben
P^I=p^I e^{\frac{p^4}{\kappa}}\,,\quad P^4=\kappa \sinh\left(\frac{p^4}{\kappa}\right)+\frac{\vec{p}^2}{2\kappa}e^{\frac{p^4}{\kappa}}\,,\quad P^5=\kappa \cosh\left(\frac{p^4}{\kappa}\right)-\frac{\vec{p}^2}{2\kappa}e^{\frac{p^4}{\kappa}}\,.
\een
Again, these cover only half of de Sitter space where $P^4+P^5>0$. The metric in these ``flat" coordinates is
\ben
ds^2=-(dp^4)^2+e^{\frac{2p_4}{\kappa}}\delta_{IJ}dp^I \, dp^J\,.
\een
Slices of constant $p_4$ are flat; to an observer using these coordinates the spacetime appears as expanding exponentially. An illuminating discussion of different coordinate systems and kinematics on de Sitter space is given in \cite{special}.

The Magueijo-Smolin model \cite{msmodel} corresponds to the following choice of coordinates:
\ben
p^1=\kappa\frac{P^1}{P^5-P^4}\,,\;p^2=\kappa\frac{P^2}{P^5-P^4}\,,\;p^3=\kappa\frac{P^3}{P^5-P^4}\,,\;p^4=\kappa\frac{P^4}{P^5-P^4}\,,
\een
The generators of boosts in de Sitter space take the form
\ben
K^I\equiv p^I\frac{\partial}{\partial p^4}+p^4\frac{\partial}{\partial p^I}+\frac{1}{\kappa}p^I p^J\frac{\partial}{\partial p^J}\,,
\een
and translations (not considered by the authors) would take the form
\ben
X_I=\frac{p^4+\kappa}{\kappa}\frac{\partial}{\partial p^I}+\frac{1}{\kappa^2} p_I p^b\frac{\partial}{\partial p^b}\,,\quad X_4=\frac{1}{\kappa}p^b\frac{\partial}{\partial p^b}+\frac{p^4+\kappa}{\kappa}\frac{\partial}{\partial p^4}\,.
\een
This choice of coordinates is somewhat peculiar as $p^4$ takes a special role, as is also apparent from the modified dispersion relations presented in \cite{msmodel}. The quantity
\ben
||p||^2=\frac{\eta_{ab}p^a p^b}{(1+\frac{1}{\kappa}p^4)^2}
\een
is invariant under boosts and rotations in de Sitter space, as would $\eta_{ab}p^a p^b$ be in Beltrami coordinates.

Each DSR model corresponds to a choice of coordinates on de Sitter space, such that all expressions reproduce the expressions for special-relativistic Minkowski coordinates as $\kappa\rightarrow\infty$. What Smolin and Magueijo call a ``$U$ map" is essentially a coordinate transformation from Beltrami coordinates to a different set of coordinates, which becomes the identity as $\kappa\rightarrow\infty$. In the remaining sections we shall use Beltrami coordinates. Note that this means we always have $p\cdot p\ge -\kappa^2$.

\setcounter{equation}{0}
\section{A de Sitter Gauge Theory of Gravity}
\label{gauge}

The most direct implementation of the ideas discussed so far into a framework describing more general spacetimes is replacing the cotangent (or tangent) bundle usually taken as phase space by a general symplectic manifold $\{\P,\omega\}$, which can be locally viewed as a product $U \times \dS$ of a subset $U\subset \M$ of spacetime $\M$ with de Sitter space $\dS$. We want to retain the differentiable structure of a manifold, which we do not assume to be present in a full theory of quantum gravity. We also assume that the structure of momentum space is fixed and in particular does not depend on matter fields, as suggested in \cite{moffat}.

If phase space is described as such a manifold, with a choice of origin in the ``tangent" de Sitter space at each point, the appropriate mathematical language is that of fibre bundles. The theory of connections on fibre bundles of this type, called {\it homogeneous bundles} in \cite{russians}, was developed by \'Elie Cartan (e.g. in \cite{cartan}). Adopting this framework means there is now an $\frak{so}(d,1)$ connection, instead of an $\frak{so}(d-1,1)$ connection, defining parallel transport on spacetime.

It was noted by MacDowell and Mansouri \cite{mmgravity} that gravity with a cosmological term in four dimensions could be obtained from a theory of such an $\frak{so}(d,1)$ connection by projecting it onto its $\frak{so}(d-1,1)$ part in the action. A more elaborate description in terms of Einstein-Cartan theory was then given by Stelle and West \cite{stellewest}. Their analysis included the gauge field needed to identify the fibres at different spacetime points, which will be crucial for the interpretation of the theory. The mathematical side of MacDowell-Mansouri gravity as a theory of a Cartan connection is nicely illustrated in \cite{wise}; we follow this article as well as the more computationally based presentation of \cite{stellewest}, who use the language of non-linear realizations. An overview over the mathematics of Cartan connections is given in \cite{sharpe}. 

For clarity we first describe the framework in a language more common to physicists; a more mathematical account of Cartan connections on homogeneous bundles is given in appendix \ref{app}. 

The usual description of general relativity as a gauge theory of the Lorentz group is known as vier-/vielbein formalism, method of moving frames, etc. Since the tangent bundle is in our description replaced by a homogeneous bundle with a curved ``tangent" space, one effectively uses a ``double vielbein" formalism, in which spacetime vectors are mapped to vectors in the tangent space to the internal (curved) space by a soldering form (vielbein). The picture we have in mind is that of a de Sitter space rolled along the manifold. One then needs to introduce a new field which specifies the point of tangency, expressed in a given coordinate system on the internal space, at each spacetime point. We denote it by $p^a(x)$. This corresponds mathematically to a necessary choice of zero section (see appendix), and physically to a gauge field. Picking a point of tangency at each spacetime point breaks the gauge group $SO(d,1)$ down to the Lorentz subgroup $SO(d-1,1)$ leaving this point invariant.

Since we consider a theory with gauge group $SO(d,1)$, the connection $A$ takes values in the Lie algebra $\frak{so}(d,1)$. It can be split as (introducing a length $l$ on dimensional grounds)
\ben
A=\left(\matrix{&  & & \cr & {\omega^a}_b & & \frac{1}{l}e^i \cr & & & \cr  & -\frac{1}{l}e_i & & 0}\right)\,,
\een
so that ${\omega^a}_b$ acts as the usual $\frak{so}(d-1,1)$-valued connection of general relativity and $e^i$ as a vielbein one-form. In doing this we have simultaneously unified the usual connection and the vielbein, and replaced the (flat) tangent space by a curved ``internal" space, such that the de Sitter group and not the Poincar\'e group now appears as a gauge group. (Lorentz) indices on ${\omega^a}_b$ and $e^i$ are now raised and lowered using $\eta^{ab}$. 

A gauge transformation, i.e. a local transformation $g(x)$ taking values in the de Sitter group, can be split as $g(x)=s(x)\Lambda(x)$, where $s(x)$ changes the zero section, i.e. changes the local identification of points of tangency at each spacetime point, and $\Lambda(x)$ is a usual local Lorentz transformation in the vielbein formalism of general relativity which does not mix the ${\omega^a}_b$ and $e^i$ parts of the connection. The connection transforms under a gauge transformation as
\ben
A(x)\rightarrow A'(x)=\Lambda^{-1}(x)s^{-1} (x) A(x) s(x)\Lambda(x) + \Lambda^{-1}(x)s^{-1} (x) ds(x)\Lambda(x) + \Lambda^{-1}(x)d\Lambda(x)\,.
\label{gaugetransf}
\een

One can use this equation to relate the connection $A_0$ corresponding to the trivial zero section, where the point of tangency is the origin of the internal space at each spacetime point, $p^a(x)\equiv (0,0,0,0)$, to a connection corresponding to any given zero section. The physical significance of this is the following. Assume we have fixed $p^a(x)\equiv (0,0,0,0)$. Then an action can be defined from the curvature of the connection $A$ (here $R$ is the curvature of the $\frak{so}(d-1,1)$ part of $A$),
\ben
F=dA+A\wedge A=\left(\matrix{& & & \cr & {R^a}_b - \frac{1}{l^2}(e^a \wedge e_b) & & \frac{1}{l}T^i\equiv\frac{1}{l}(de^i + {\omega^i}_j \wedge e^j) \cr \cr & -\frac{1}{l}T_i & & 0}\right)\,.
\een
In four dimensions, the MacDowell-Mansouri action \cite{mmgravity, wise} is
\ben
S=-\frac{3}{32\pi G\Lambda}\int \epsilon_{abcd} \left(F^{ab} \wedge F^{cd}\right)= -\frac{3}{32\pi G\Lambda}\int d^4 x \, \frac{1}{4}\epsilon_{abcd}\epsilon^{\mu\nu\rho\tau}F_{\mu\nu}^{ab}F_{\rho\tau}^{cd}\,,
\label{akshn}
\een
where the Latin indices run from 1 to 4, and so one projects $F$ to its $\frak{so}(d-1,1)$ part in this action. 

Apart from a topological Gauss-Bonnet term, the action (\ref{akshn}) is equivalent to the Einstein-Hilbert action with a cosmological term
\ben
S = \frac{3}{16\pi G\Lambda}\frac{1}{l^2}\int \epsilon_{abcd} \left(e^a\wedge e^b \wedge R^{cd} - \frac{1}{2 l^2}e^a\wedge e^b \wedge e^c \wedge e^d \right)\,,
\label{einsthilb}
\een
where we have to identify
\ben
\Lambda=\frac{3}{l^2}\,.
\een
as the cosmological constant.

In order to define the projection of $F$ in the action (\ref{akshn}), one has used a splitting 
\ben
\frak{so}(d,1) \simeq \frak{so}(d,1)/\frak{so}(d-1,1) \oplus \frak{so}(d-1,1),
\label{split1}
\een
 which depends on the gauge field since the subgroup $SO(d-1,1)$ leaving a given point in de Sitter space invariant depends on the choice of this point. 

When the action (\ref{akshn}) is coupled to matter, the $\frak{so}(d,1)/\frak{so}(d-1,1)$ part $e^a$ of the connection appears in a volume element in the matter Lagrangian. By varying the action one obtains the field equations of Einstein-Cartan theory with a cosmological constant $\Lambda=3/l^2$. The length scale $l$, which is so far arbitrary, can be chosen to reproduce the $\Lambda$ of the observed universe, which means it must be chosen to be very large (the ``cosmological constant problem"). By the field equations, one can determine for a given matter distribution a connection $A_0$ consisting of an $\frak{so}(d-1,1)$ connection $({\omega^a}_b)_0$ and a vielbein $e^i_0$. 

The MacDowell-Mansouri action reproducing Einstein-Cartan theory with a cosmological constant includes a gauge choice. We can hence view it as the gauge-fixed version of a more general theory. Since (\ref{gaugetransf}) determines how the connection transforms under a gauge transformation, we can generalise a given solution of Einstein-Cartan theory to an arbitrary gauge choice. The extension of the theory to arbitrary configurations of the gauge field, and hence arbitrary choices of tangency points of the internal space to spacetime, is what we call {\it Einstein-Cartan-Stelle-West theory}. Any solution of Einstein-Cartan theory, in particular any (torsion-free) solution of general relativity, gives rise to more general solutions of Einstein-Cartan-Stelle-West theory via (\ref{gaugetransf}). We will later see that one can construct an $\frak{so}(d-1,1)$ connection with torsion from a torsion-free one.

In (\ref{gaugetransf}), $s(x)$ takes values in the de Sitter group, more precisely in the subgroup generated by ``translations" which leaves no point of de Sitter space invariant. The correspondence between Beltrami coordinates $p^a(x)$ on de Sitter space and such group elements is given explicitly by
\ben
s(p(x))=\exp\left[ \frac{p^i(x)}{\sqrt{-p(x)\cdot p(x)}}\,{\rm Artanh}\left(\frac{\sqrt{-p(x)\cdot p(x)}}{\kappa}\right)\kappa\,X_i \right]\,.
\label{param}
\een 
Then the group element $s(p(x))$ maps $(0,0,0,0)$ to $(p^1(x),p^2(x),p^3(x),p^4(x))$ in Beltrami coordinates. A different choice of coordinates in the internal de Sitter space would correspond to a different parametrisation of the elements of the subgroup of translations of the de Sitter group.

Inserting (\ref{param}) into (\ref{gaugetransf}) and setting $\Lambda(x)\equiv e$, we obtain
\bena
\omega^{ab}(p(x)) & = & \frac{p^a e_0^b}{l \kappa \gamma(p)} + \left(1 - \frac{1}{\gamma(p)}\right)\frac{p^a dp^b + \omega_0^{ca}p^b p_c}{p\cdot p} + \frac{1}{2}\omega_0^{ab} - (a \leftrightarrow b)\,,
\label{explicit}
\\ e^i(p(x)) & = & \frac{l \kappa}{p\cdot p+\kappa^2}\left(p^i \frac{p_c dp^c }{p\cdot p}(1-\gamma(p)) +dp^i \gamma(p) + ({\omega^i}_b)_0 p^b \gamma(p)\right)+p^i e_0^a p_a\frac{1 + \frac{\kappa^2}{p\cdot p}(1-\gamma(p))}{p\cdot p+\kappa^2} + \frac{e_0^i}{\gamma(p)}\,, \nn
\eena
where
\ben
\gamma(p)\equiv\sqrt{\frac{p\cdot p +\kappa^2}{\kappa^2}}=1+\frac{p\cdot p}{2\kappa^2}+\ldots\,
\een

Because $p\cdot p \ge -\kappa^2$ in Beltrami coordinates, the square root is always real. In the limit $p\cdot p\rightarrow 0$, our parametrisation is the same as that used in \cite{stellewest}, and we recover their results
\bena
 \omega^{ab}(p(x)) & = & \left(\half\omega^{ab}_0+\frac{1}{l \kappa} p^a e_0^b + \frac{1}{2\kappa^2} \left(p^a dp^b + \omega_0^{ca}p^b p_c \right)\right) - (a \leftrightarrow b)\,,
\\ e^i(p(x))& = & e_0^i + \frac{l}{\kappa}\left(-\frac{1}{2\kappa^2}p^i p_c dp^c +dp^i + \omega^{ib}_0 p_b \right)+\frac{1}{2\kappa^2} p^i e_0^a p_a \,. \nn
\eena
Near $p=0$, we have
\ben
\omega^{ab}(p(x)) = \omega_0^{ab} + O\left(\frac{p}{\kappa}\right)\,, \quad e^i(p(x)) = e_0^i + \frac{l}{\kappa}dp^i + O\left(\frac{p}{\kappa}\right)\,. \nn
\label{smallp}
\een

As mentioned above, the $\frak{so}(d,1)/\frak{so}(d-1,1)$ part of the connection $A$ acts as a vielbein and maps vectors in the tangent space at a point $x$ in spacetime to vectors in the tangent space at $p(x)$ in the internal de Sitter space, given in components with respect to an orthonormal basis at $p(x)$. In order to give their components in the coordinate-induced basis $\{\frac{\partial}{\partial p^a}\}$, we need another vielbein, which can be obtained from (\ref{explicit}) by setting $\omega_0=e_0=0$ (corresponding to spacetime being de Sitter space with cosmological constant $\Lambda$) and $p^a(x)=\frac{\kappa}{l}x^a$, as in \cite{stellewest}. We obtain
\ben
{l_n}^a(p(x))=\kappa^2\frac{{\delta_n}^a (p\cdot p) \gamma(p) - p^a p_n (\gamma(p)-1)}{(p \cdot p)(p\cdot p+\kappa^2)}\,,
\label{vierbein}
\een
where $n$ is a coordinate index in the internal space and $a$ denotes a Lorentz index, as before. This vielbein is of course independent of the underlying spacetime.

Parallel transport can be defined for the $\frak{so}(d,1)$ connection using the notion of development, which generalises the usual covariant derivative. One introduces a development operator \cite{stellewest}
\ben
D = d - \half\omega^{ab}M_{ab} - (e\cdot V)\,,
\een
where the second term is the usual infinitesimal relative rotation of tangent spaces at different spacetime points, and the last term compensates for the change of point of tangency and hence generates maps from the tangent space at one point of de Sitter space to the tangent space at a different point of de Sitter space. Again one should think of an internal space rolled along spacetime \cite{wise}. 

In components, in our conventions we have
\ben
{(\omega^{ab}M_{ab})^c}_d=-2{\omega^c}_d\,,
\een
and the combination $e^a V_a$ acts on Lorentz indices as an element of $\frak{so}(d-1,1)$, representing the map from one tangent space to another in the respective bases. We use the result obtained by \cite{stellewest} using the techniques of non-linear realizations\footnote{For an exposition of the theory of non-linear realizations, see \cite{coleman}.}, namely that when expressed as an $\frak{so}(d-1,1)$ matrix,
\ben
l(e\cdot V) = \kappa\,s(p)^{-1}(e^a X_a )s(p)-s(p)^{-1}\left[s(p+\delta p)-s(p)\right]\,,
\een
where $s(p)$ is defined according to (\ref{param}) and $\delta p$ is determined from the equation
\ben
{\left[s(p+\delta p)\right]^a}_5={\left[(1+e^b X_b \kappa)s(p)\right]^a}_5
\een
where only terms linear in $e^a$ are kept in $\delta p$. An explicit calculation shows that
\ben
\delta p^a = \frac{p^a}{\kappa}(\eta_{bc}e^b p^c)+e^a\kappa\,,
\een
and hence near $p=0$, we have $\delta p^a = \kappa e^a$, as expected. We find that $(e\cdot V)$ has components
\ben
{(e\cdot V)^b}_c=\frac{\kappa(e^b p_c - e_c p^b)(1-\gamma(p))}{l (p\cdot p)}\,.
\label{development}
\een

\begin{figure}[h]
\begin{picture}(300,120)
\put(0,0){\line(1,0){200}}\put(0,0){\line(1,2){50}}
\put(50,100){\line(1,0){200}}\put(200,0){\line(1,2){50}}
\put(100,60){\circle{40}}\bezier{313}(80,60)(100,50)(120,60)
\bezier{125}(100,40)(115,40)(120,30)\bezier{578}(100,40)(115,20)(120,30)
\bezier{125}(100,40)(110,42)(112,50)\bezier{578}(100,45)(110,55)(112,50)
\end{picture}
\caption{When the curved internal space is rolled along Minkowski space, a path in spacetime corresponds to a path in the internal space. Because of the curvature of the internal space, a closed path in Minkowski space does not correspond to a closed path in the internal space, which is manifest as torsion.}
\label{fig}
\end{figure}
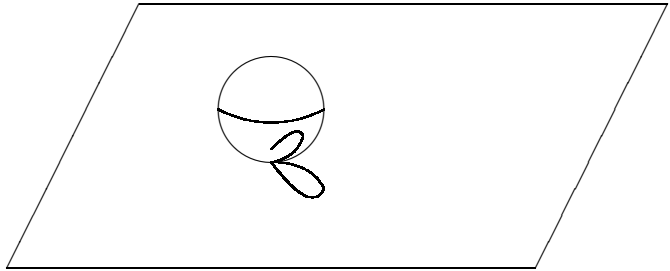

One then has a notion of holonomy, mapping closed loops in spacetime into the internal space by development. In particular, if one develops the field $p(x)$ describing the point of tangency around an infinitesimal closed loop at $x_0$, the developed value will in general differ from the original value at $x_0$ \cite{stellewest}:
\ben
\Delta p^a (x_0) \propto {T_{\mu\nu}}^i(x_0){l^a}_i(p(x_0))\oint x^{\mu} dx^{\nu}\,,
\een
where ${l^a}_i(p(x))$ is the inverse of the vielbein (\ref{vierbein}) and ${T_{\mu\nu}}^i$ are the components of the torsion tensor $T=de+\omega\wedge e$. The situation for Minkowski space, which we will discuss next, is illustrated in figure \ref{fig}. The central result we will try to justify in the following is that, starting from Minkowski spacetime, if we assume the internal space is rolled along Minkowski space in a non-trivial way, we obtain a connection with torsion. In our interpretation, this is the only way that ``coordinates" can act as translations on momentum space, as one normally assumes when associating the Snyder algebra with a noncommutative spacetime.

\setcounter{equation}{0}
\section{Synthesis}
\label{synth}

The notion of development along curves in spacetime is central to the interpretation of Einstein-Cartan-Stelle-West theory, because it allows ``spacetime coordinates" to act as translations in the internal de Sitter space. The situation described by DSR, where noncommuting translations on a curved momentum space are interpreted as noncommuting spacetime coordinates, here corresponds to a Minkowski spacetime with an internal de Sitter space rolled along this Minkowski space. The gauge field $p^a(x)$ specifies the points of tangency of the internal space at each spacetime point, and we have chosen Beltrami coordinates on de Sitter space which look like Cartesian coordinates on Minkowski space near the ``origin" of de Sitter space. Since the internal space has a natural scale $\kappa$ and we needed to introduce a natural scale $l$ in spacetime, we choose the gauge field to be
\ben
p^a(x)=\frac{\kappa}{l}x^a
\label{choice}
\een
in a vicinity of the origin of spacetime which is now taken to be Minkowski space, where $x^a$ are the standard Minkowski coordinates such that the connection vanishes in general relativity. In general a closed path in spacetime will not correspond to a closed path traced out on the internal space, hence such an identification is only local and, strictly speaking, only valid the origin of Minkowski spacetime. On dimensional grounds, the effects of torsion scale as $\frac{x}{l}$ or $\frac{p}{\kappa}$. For (\ref{choice}) to be well-defined, we must guarantee that $x\cdot x \ge -l^2$, so $l$ should be large in Planck units. We will comment on the significance of the scale $l$ at the end of this section.

It should perhaps be emphasised that the gauge field $p^a(x)$ does not represent physical momentum, but determines the point of tangency of the internal space we have introduced which is to some extent arbitrary. Tangent vectors to the original spacetime can be mapped to tangent vectors to the internal space via the vielbein. The physical interpretation of motion in an internal ``momentum" space which is related to motion in spacetime seems obscure, but if coordinates are to act as translations in the internal space, the two must be connected in some way. In this sense, we are constructing the minimal non-trivial gauge field which leads to observable effects, and an alternative interpretation of noncommuting generators $X_a$ in the Snyder algebra. 

Since we do not interpret different points in the internal de Sitter space as representing different values for physical four-momentum, we avoid problems with the physical interpretation of DSR, such as the ``spectator problem" of noncommutative momentum addition and the ``soccer ball problem" of how to describe extended objects. In our framework, tangent vectors representing a particle's (or extended body's) velocity remain vectors and as such live in an unbounded space with commutative addition.

As explained before, we can use equations (\ref{explicit}) to obtain the connection components $\omega$ and $e$ corresponding to this choice of our gauge field; we set $\omega_0=0$ and $({e_{\mu}}^a)_0={\delta_{\mu}}^a$ and substitute (\ref{choice}) to get
\bena
{\omega_{\mu}}^{ab} & = & \left(x^a {\delta_{\mu}}^b - x^b {\delta_{\mu}}^a \right)\frac{x\cdot x+l^2(\gamma(p)-1)}{l^2 (x\cdot x) \gamma(p)}\,,
\label{connection}
\\{e_{\mu}}^i & = & \frac{1}{(x\cdot x)(x\cdot x + l^2)}\Big(x^i x_{\mu}(x\cdot x-2 l^2(\gamma(p)-1))+{\delta_{\mu}}^i (x\cdot x) 2 l^2 \gamma(p)\Big) \nn
\eena
and
\bena
\partial_{\nu}{e_{\mu}}^i-\partial_{\mu}{e_{\nu}}^i & = & \left(x_{\nu}{\delta_{\mu}}^i-x_{\mu}{\delta_{\nu}}^i\right)\frac{l^2(2l^2(\gamma(p)-1)-3(x\cdot x))-(x\cdot x)^2}{(x\cdot x)(x\cdot x+l^2)^2}\,,\nn
\\{\omega_{\nu}}^{ib}e_{\mu b}-{\omega_{\mu}}^{ib}e_{\nu b} & = & \frac{x\cdot x+ l^2(\gamma(p)-1)}{(x\cdot x)(x\cdot x+l^2)l^2\gamma(p)}\left(x_{\nu}{\delta_{\mu}}^i-x_{\mu}{\delta_{\nu}}^i\right)\left(2 l^2+x\cdot x\right)\,,
\eena
which gives a non-zero torsion
\ben
{T_{\mu\nu}}^i=\left(x_{\nu}{\delta_{\mu}}^i-x_{\mu}{\delta_{\nu}}^i\right)\frac{1}{l^2 \sqrt{\frac{x\cdot x}{l^2}+1}}\,.
\een
Interestingly enough, for the choice of zero section (\ref{choice}) the scale $\kappa$ drops out of all expressions. Expressed in coordinates on the internal space, one has
\ben
{T_{\mu\nu}}^i=\left(p_{\nu}{\delta_{\mu}}^i-p_{\mu}{\delta_{\nu}}^i\right)\frac{1}{l \kappa \sqrt{\frac{p\cdot p}{\kappa^2}+1}}\,.
\een
The quantity ${T_{\mu\nu}}^i$ will be multiplied by an infinitesimal closed loop $\oint x^{\mu} dx^{\nu}$ to give the difference in the value $p(x)$ caused by development along this loop. In momentum coordinates, this is equal to $\frac{l}{\kappa}\oint p^{\mu} dp^{\nu}$, and the effect of going around the developed curve in the internal space is (near $x=0$ or $p=0$) proportional to $\kappa^{-2}$, just as was suggested by (\ref{algebra}). 

Expressing Minkowski space in the usual coordinates, together with the (local) identification $p^a(x)=\frac{\kappa}{l}x^a$, in this framework gives a connection with torsion. Developing a closed curve in spacetime in the internal space will give a curve that does not close in general, which is the effect of noncommuting translations in the internal space. 

The reader may wonder how the ``deformation" of the Minkowski solution described here is manifest in a metric. We can define a metric by the usual expression
\ben
g_{\mu\nu}=e_{\mu}^a e_{\nu}^b \eta_{ab}\,.
\een
This metric would not determine the connection, but could be used to define distances in the spacetime in the usual way. Then, from (\ref{connection}), we get
\ben
g_{\mu\nu}=\eta_{\mu\nu}\,\frac{4}{1+\frac{x \cdot x}{l^2}}+x_{\mu}x_{\nu}\frac{(x\cdot x)}{\left((x\cdot x)+l^2\right)^2}\,.
\een
It should be stressed that the connection on spacetime is {\it not} the Levi-Civita connection of this metric. There is a factor of 4 because of a term in (\ref{smallp}) which does not necessarily go to zero as $p\rightarrow 0$. With the identification (\ref{choice}), the soldering form always gets a contribution
\ben
{e_{\mu}}^i(x) = ({e_{\mu}}^i)_0 + {\delta_{\mu}}^i + O\left(\frac{x}{l}\right)\,.
\een
The limit $\kappa\rightarrow\infty$ is now identified with the limit $l\rightarrow\infty$, in which we recover the (rescaled) Minkowski metric. 

In deriving the expressions (\ref{connection}) we started with Minkowski space, which clearly solves the field equations of the Einstein-Cartan theory for an energy-momentum tensor cancelling the cosmological constant term, and vanishing internal spin. In changing the zero section, we then performed a $\dSG$ gauge transformation, under which the curvature $F$ transformed as
\ben
F(s(x))=s^{-1} (x) F(x) s(x).
\een
Since this is a general $\dSG$ rotation, it mixes up the $\frak{so}(d-1,1)$ and $\frak{so}(d,1)/\frak{so}(d-1,1)$ parts of the connection and the curvature. Hence, the resulting connection will no longer solve the original field equations, but the field equations for an energy-momentum tensor which has also undergone a $\dSG$ transformation. This mixes up the energy-momentum and internal spin parts, combining them into an element of the Lie algebra $\frak{so}(d,1)$, the interpretation of which seems obscure at least.

A comment is in order with regard to physical units. In addition to the energy scale $\kappa$, which is perhaps naturally identified with the Planck scale, the identification of lengths with momenta, necessary in the framework presented here, requires the choice of a unit of length $l$ which is not necessarily connected to the scale $\kappa$. It may well be that it is instead the cosmological constant which sets this length scale, leading to an astronomical scale instead of a sub-atomic one. And indeed, some more recent approaches to quantum gravity (e.g. \cite{friedel, ita}) use the product ${G\Lambda}$ as a dimensionless parameter in a perturbative expansion. A fixed positive $\Lambda$ is also required in non-perturbative approaches to quantum gravity \cite{quantsym}. Then the cosmological constant may play the role of a fundamental parameter in quantum gravity.

\setcounter{equation}{0}
\section{Discussion}

It has been argued that the algebra of DSR describes the symmetries of a semiclassical limit of (a generic theory of) quantum gravity (see e.g. \cite{quantsym}). If this claim is taken seriously, one has to give an interpretation of the noncommuting translations appearing in the algebra, and usually they are supposed to represent a spacetime with a fundamentally noncommutative structure \cite{madore}. Alternatively, one may view the apparent noncommutativity as an artefact of the finite resolution of lengths \cite{oriti}. However, there are fundamental difficulties in associating these operators directly with coordinates on spacetime, as position is not {\it additive} in a way that momentum and angular momentum are \cite{stability}. Furthermore, as also pointed out in \cite{stability}, a proposed noncommutativity of spacetime of the form (\ref{algebra}), proportional to angular momentum or boost generators, and hence vanishing at a given ``origin", seems deeply at odds with any idea of (even Galilean) relativity. This would also be an obvious criticism of the framework presented in this note, when taken as a theory that is supposed to describe the real world.

What we have shown here, is that using the framework of Einstein-Cartan-Stelle-West theory, one reaches a different conclusion from the usual one: The noncommutativity of translations on a momentum space of constant curvature is interpreted as torsion of a connection that solves the equations of Einstein-Cartan theory with a modified energy-momentum tensor that mixes with the spin tensor. If one takes this seriously, one is led to conclude that there is an effect of torsion induced by quantum gravity, whose effects would however only become measurable over distances comparable to $l$, a length scale presumably associated with the cosmological constant.

No such effect appears in de Sitter space with an appropriate cosmological constant, or indeed any vacuum solution of the theory. Vacuum solutions are then just described by the Poincar\'e algebra, and hence undeformed special relativity.

Any non-zero energy-momentum tensor, however, will lead to a connection having torsion. In theories such as Einstein-Cartan theory, this leads to well-known problems when trying to couple the gravitational field to Maxwell fields, for instance, as there is no unambiguous procedure of minimal coupling. This is because the statement that the exterior derivative is independent of the choice of connection,
\ben
(dA)_{\mu\nu}\propto\partial_{[\mu}A_{\nu]}=\nabla_{[\mu}A_{\nu]}\,,
\een
is true precisely when torsion vanishes. Using an $\frak{so}(d-1,1)$ connection, this is apparent from
\ben
d(e^i A_i)=\nabla(e^i A_i)=A_i \nabla e^i - e^i \wedge\nabla A_i = - e^i \wedge \nabla A_i + A_i T^i
\een
where $\nabla e^i=de^i+{\omega^i}_j\wedge e^j$ etc. One has two different candidates for the field strength $F$, namely $e^i\wedge \nabla A_i$ and $d(e^i A_i)$, with possibly observable differences between these choices, although it could be argued that $F=dA$ is the only meaningful choice because it preserves gauge invariance \cite{benn}.

In the framework of Einstein-Cartan-Stelle-West theory, gauge fields should be coupled to gravity via development, i.e. replacing $F=dA$ by $F=DA$. We compute from (\ref{explicit}) and (\ref{development}) that development can be expressed in terms of $\omega_0$ and $e_0$ by
\ben
D=d - \half\omega^{ab}M_{ab} - (e\cdot V) = d + \omega - (e\cdot V) = d + \omega_0 + 2(p \otimes_A e_0)\kappa\frac{(\gamma(p)-1)}{l(p\cdot p)}=: d + \omega_{{\rm eff}}\,,
\een
where $\otimes_A$ is an antisymmetrised tensor product, $2(U\otimes_A V)^{ab} = U^a V^b - U^b V^a$. Parallel transport is effectively described by the connection $\omega_{{\rm eff}}$, whose torsion is in general non-zero. One can give an explicit formula for the torsion which is however rather complicated and does not seem to give much insight; to linear order in $p^i$, one has
\ben
T^i = \frac{l}{\kappa}({R^i}_b)_0 p^b - \frac{1}{2 l\kappa}e^i_0 \wedge (e^0_j p^j) + \frac{1}{2\kappa^2}\left(p^i e_{j0}\wedge dp^j - p_j e_0^i \wedge dp^j\right)+O(p^2).
\een
If we assume a universal relation of internal momenta and spacetime lengths of the form $p\sim\frac{\kappa}{l}x$, the second and third terms seem to give contributions of order $x/l^2$. The first term is proportional to the local curvature of $\omega_0$, ${R^i}_b = d{\omega^i}_b + {\omega^i}_j\wedge{\omega^j}_b$, contracted with $x^b$. Note that it is the Riemann tensor, not the Ricci tensor, that appears, so that propagating degrees of freedom of the gravitational field are included. This first term should in realistic situations, even in vacuum, give the dominant contribution.

Assuming that minimal coupling is achieved through the development operator $D$, or equivalently by using the effective connection which has torsion, one would couple vector or matter fields (using $D\psi$ for spinors) to torsion, breaking gauge invariance. Such an effect of course leads to the absence of charge conservation, and this should be experimentally observable in the presence of a non-trivial gravitational field, i.e. in regions where spacetime is not exactly de Sitter. Let us recall that in standard tensor calculus one uses the identity
\ben
[\nabla_{\mu},\nabla_{\nu}]M_{\lambda\rho}={R_{\mu\nu\lambda}}^{\sigma}M_{\sigma\rho}+{R_{\mu\nu\rho}}^{\sigma}M_{\sigma\lambda}-{{T_{\mu}}^{\sigma}}_{\nu}\nabla_{\sigma}M_{\lambda\rho}
\een
which gives for an antisymmetric $M_{\lambda\rho}$ when contracted
\ben
[\nabla^{\lambda},\nabla^{\rho}]M_{\lambda\rho}=-g^{\mu\lambda}g^{\nu\rho}{{T_{\mu}}^{\sigma}}_{\nu}\nabla_{\sigma}M_{\lambda\rho},
\een
to establish that the right-hand side of Maxwell's equation $\nabla^{\lambda}F_{\lambda\rho}=4\pi J_{\rho}$ satisfies a continuity equation in the absence of torsion. With torsion present, one has then for any region $R$
\ben
\int_{\partial R}d^3x \;\sqrt{h}\;n^{\lambda} J_{\lambda} = \frac{1}{4\pi}\int_R d^4x\; \sqrt{g}\;\left(-g^{\mu\lambda}g^{\nu\rho}{{T_{\mu}}^{\sigma}}_{\nu}\nabla_{\sigma}F_{\lambda\rho}\right).
\een
Effects become important when the size of the region $R$ is comparable to the length scale of torsion.

As an example consider the Schwarzschild solution, which has Kretschmann scalar
\ben
R_{abcd}R^{abcd} \sim \frac{r_S^2}{r^6},
\een
so roughly $R_{abcd}\sim r_S r^{-3}$. Assuming that the origin of the $x$ coordinate system corresponds to the centre of the Earth, we would, on the surface of the Earth, measure a torsion of order $r_S R^{-2}$, where $R$ is the radius of the Earth. Since $r_S\sim 10^{-2}$ m and $R^2 \sim 10^{13}\,{\rm m}^2$, this means that the length scale for effects of torsion would be about $10^{11}$ m. The other two contributions, given that $l\sim 10^{26}$ m, would be much smaller. Although this crude estimate suggests that effects will be very small, even tiny violations of charge conservation should have been observed experimentally. For a discussion of experimental tests of charge conservation and possible extensions of Maxwell theory in Minkowski space, see \cite{exptest}. Processes such as electron decay on a length scale of $10^{11}$ m, or a time scale of $10^{3}$ s, can clearly be ruled out. 

The example presented here shows that the correct physical interpretation of purely algebraic relations, such as the commutators of the Snyder algebra, may not be the seemingly obvious one. We conclude that the physical motivation for assuming spacetime is ``noncommutative" may not be as clear as often assumed. 

\section{Gauge Invariance Broken?}

The idea that an asymmetry between the proton and electron charges could have interesting astrophysical consequences goes back to Lyttleton and Bondi \cite{bondi}, who argued that a charge difference, and hence a net charge of the hydrogen atom, of $10^{-18}$ elementary charges, might explain the observed expansion of the universe by electrostatic repulsion. This idea was proposed in connection with Hoyle's ideas of a universe in a steady state, which required continuous production of material via a ``creation field" \cite{hoyle}, and a modification of Maxwell's equations was proposed to accommodate charge nonconservation. From Hoyle's perspective, however, the steady state model was incompatible with expansion of the universe by electrostatic repulsion, and should lead to electrostatic attraction instead \cite{hoyle2}. 

There seems to be need for the electron and proton charges to be of equal magnitude to maintain gauge invariance. However, if the universe as a whole is not neutral, but it is homogeneous, gauge invariance must be broken. Hence the two issues are closely related. Modern laboratory experiments \cite{laboratorium} give a bound of $10^{-21}$ elementary charges on the difference of electron and proton charge; astrophysical considerations give bounds of $10^{-26}$ elementary charges using the isotropy of the cosmic microwave background \cite{cmbbounds}, or $10^{-29}$ elementary charges by considering cosmic rays \cite{raybounds}. Recently, an interesting proposal to measure net charges of atoms and neutrons, sensitive to $10^{-28}$ elementary charges, was put forward \cite{expproposal}.

From a theoretical viewpoint, if gauge invariance is broken, it is natural to assume a nonvanishing photon mass. One then considers Einstein-Proca theory, an outline of which can be found in \cite{hejna}. The photon may also be charged. Here, experimental bounds on the charge are $10^{-29}$ elementary charges using pulsars \cite{pulsar}, and possibly $10^{-35}$ elementary charges from CMB isotropy \cite{cmbbounds}. 

Experimental bounds on violations of gauge invariance in electrodynamics are very tight, and hence any theory predicting torsion which is coupled to electromagnetism faces severe problems when confronted by experiment. In the framework of Einstein-Cartan-Stelle-West theory, it is possible to maintain gauge invariance by choosing $F=dA$, but using the development operator is the most natural choice.

\appendix
\renewcommand{\theequation}{A.\arabic{equation}}
\setcounter{equation}{0}
\section{Symmetric Affine Theory}

If Einstein-Cartan theory is considered as the extension of general relativity which allows for torsion, there is an analogous extension which allows for a non-metric connection. This theory can be formulated in terms of a torsion-free $\frak{gl}(n,\bR)$ connection and is known as {\it symmetric affine theory}. It is equivalent to standard general relativity with a massive vector field, known as (nonlinear) Einstein-Proca theory \cite{hejna}.

One could attempt to embed this theory into a theory of a connection taking values in the algebra of the affine group $\frak{a}(n,\bR)$\footnote{For a comprehensive review of general theories of this type, see \cite{hehl}.},
\ben
A=\left(\matrix{&  & & \cr & {\omega^a}_b & & \frac{1}{l}e^i \cr & & & \cr  & 0 & & 0}\right)\,,
\een
where now ${\omega^a}_b$ is not constrained by $\omega^{ab}=-\omega^{ba}$. Geometrically, this means that the connection does not preserve the lengths of vectors under parallel transport.

The corresponding curvature of $A$ ($R$ is the curvature of the $\frak{gl}(n,\bR)$ part),
\ben
F=dA+A\wedge A=\left(\matrix{& & & \cr & {R^a}_b & & \frac{1}{l}T^i\equiv\frac{1}{l}(de^i + {\omega^i}_j \wedge e^j) \cr \cr & 0 & & 0}\right)\,,
\een
would then be constrained by demanding that $T^i\equiv 0$. This seems rather unnatural from the perspective of Cartan geometry. Furthermore, the length scale $l$ is now completely arbitrary as it does not appear in the $\frak{gl}(n,\bR)$ part of the curvature any more.

One proceeds by considering Lagrangians that only depend on the Ricci tensor, which is a one-form obtained by contracting the components of the Riemann curvature, written in the basis of one-forms given by the vielbein $e^i$:
\ben
{\rm Ric}_a = {\rm Ric}_{ia}e^i,\quad {\rm Ric}_{ia}={{R_{ji}}^j}_a,
\een 
where the curvature two-form is
\ben
{R^a}_b = \frac{1}{2}{{R_{ij}}^a}_b e^i\wedge e^j.
\een
One then splits the Ricci tensor into symmetric and antisymmetric part, symmetrising over a component (with respect to the given basis) index and a $\frak{gl}(n,\bR)$ index. The antisymmetric part can be interpreted as a spacetime two-form
\ben
i_{e_a}\left({R^a}_b\wedge e^b\right),
\een
where $i_{e_a}$ is interior multiplication with the vector ${e_a}$, defined by being dual to the one-forms ${e^b}$:
\ben
e^b(e_a)={\delta^b}_a.
\een
No such construction is possible for the symmetric part, which is normally more relevant in concrete constructions. The splitting itself seems depend on the choice of basis.

\renewcommand{\theequation}{B.\arabic{equation}}
\setcounter{equation}{0}
\section{Cartan Connections on Homogeneous Bundles}
\label{app}

This more mathematical introduction into Cartan connections on homogeneous bundles relies mainly on \cite{wise}, but mentions some additional points which are of importance to our discussion of Einstein-Cartan-Stelle-West theory.

The tangent bundle of a manifold needs to be replaced by a fibre bundle whose fibres are homogeneous spaces $\dS\equiv\dSG/SO(d-1,1)$. This can be achieved by starting with a principal bundle $P(\M,\dSG)$, and considering the associated bundle $\P=E(\M,\dS,\dSG,P)=P \times_{\dSG} \dS$ (taken as phase space); it can be identified with $P/SO(d-1,1)$ by the map
\ben
\nu: \P \rightarrow P/SO(d-1,1), \quad [u,a\cdot SO(d-1,1)] \mapsto u a \cdot SO(d-1,1).
\een
Then the structure group $\dSG$ is reducible to $ SO(d-1,1)$ if the associated bundle $\P$ admits a cross section $\sigma: \M \rightarrow \P$ \cite{kobayashi}; furthermore, there is a one-to-one correspondence between reductions of the structure group and cross sections. This cross section, called a {\it zero section} in \cite{petti}, corresponds to a choice of origin in the momentum space attached to each point. In physicist's terms, the de Sitter group is spontaneously broken down to the Lorentz group by the choice of points of tangency in the tangent de Sitter spaces at each spacetime point.

The bundle reduction depends on the choice of zero section, or rather, its local representation in coordinates as a function $\M \supset U \rightarrow U \times \dS$. This is because the embedding of $ SO(d-1,1)$ into $\dSG$ is not canonical, as the stabilizers of different points in $\dS$ are isomorphic but related by conjugation. In other words, the mappings appearing in the exact sequence
\ben
{\bf 0} \rightarrow  SO(d-1,1) \rightarrow \dSG \rightarrow \dSG /  SO(d-1,1) \rightarrow {\bf 0}
\een
are not canonically chosen (cf. the discussion for the affine group in \cite{petti}).

It is of course possible to choose canonical coordinates such that the function representing the zero section is just $x\mapsto (x,[e])\equiv (x, SO(d-1,1))\in \M \times \dS$. However, in general we want to locally identify the fibres at nearby base space points, adopting the viewpoint that there is a single tangent $\dS$ space which is ``rolled along" the manifold. Then we need to retain the general coordinate freedom. (This point is missing in the discussion of \cite{wise}.) An exact identification is only possible when the connection is flat. Let us assume that coordinates on $\P$ have been fixed, and that it is the zero section, and hence the identification of the fibres, that is varied\footnote{one is free to choose an ``active" or ``passive" viewpoint here}. After a choice of zero section, there is still a local gauge freedom corresponding to the stabilizer $SO(d-1,1)$. We express a given section as $s(x)$, where $\sigma(x)=(x,s(x))\in \M \times \dS$ in our coordinates. The section that corresponds to $s_0(x)\equiv [e]$ will be called ``trivial".

An $\frak{so}(d,1)$-valued Ehresmann connection $\mathbf{A}$ in $P$ is in general not reducible to an $\frak{so}(d-1,1)$-valued connection in the reduced $ SO(d-1,1)$ bundle $P_R(\M, SO(d-1,1))$. It can, however, be pulled back using the inclusion
\ben
\iota_x:  SO(d-1,1) \rightarrow \dSG, \quad  \Lambda \mapsto s(x)\Lambda s(x)^{-1}
\een
to a Cartan connection $\mathbf{A}_C$ on the reduced bundle\footnote{We assume here that the necessary condition $\ker\mathbf{A} \cap (\iota_x)_*(TP_R(\M, SO(d-1,1)))=\{0\}$ (see \cite{sharpe}) is satisfied.}. Of course reducing the connection to an $\frak{so}(d-1,1)$-valued connection and pulling it back to a Cartan connection are very different operations, since in the latter case one wants the $\frak{so}(d,1)/\frak{so}(d-1,1)$ part of the pulled-back connection to act as a soldering form, so in particular to be non-singular. We obtain a bundle sequence (cf. \cite{wise})
\begin{center}
\begin{picture}(200,110)
\put(-50,90){$P_R(\M, SO(d-1,1))$}\put(50,93){\vector(1,0){25}}\put(80,90){$P(\M,\dSG)$}\put(160,93){\vector(1,0){25}}\put(190,90){$P/\iota_x( SO(d-1,1))\simeq \P$}\put(280,100){{\small $\nu_x^{-1}$}}
\put(35,80){\vector(1,-1){60}}\put(110,80){\vector(0,-1){60}}\put(185,80){\vector(-1,-1){60}}\put(105,5){$\M$}
\end{picture}
\end{center}
The reduced bundle $P_R(\M, SO(d-1,1))$ is mapped into $P(\M,\dSG)$ by
\ben
p \mapsto [p,e]=\{(p\Lambda^{-1},s(x)\Lambda s(x)^{-1})|\Lambda \in  SO(d-1,1)\}\in P_R(\M, SO(d-1,1)) \times_{\iota_x( SO(d-1,1))} \dSG
\een
The connection one-form $A$ on $\M$, induced by the connection $\mathbf{A}$ on $P$, depends on a choice of section $\tau:\M\rightarrow P(\M,\dSG)$. If the zero section $\sigma$ is fixed, one can choose an arbitrary (local) section $\tau_R:\M\rightarrow P_R(\M, SO(d-1,1))$ to obtain a section $\tau$; in local coordinates,
\ben
\sigma(x)=(x,s(x))\,,\; \tau_R(x)=(x,\Lambda(x))\;\longrightarrow\; \tau(x)=(x,s(x)\Lambda(x) s(x)^{-1}\cdot s(x))=(x,s(x)\Lambda(x))\,.
\een
For practical computations, it is often useful to first consider the trivial section. The induced connection corresponding to this section, denoted by $A_0(x)$, is related to the connection for a general section by
\ben
A(\tau(x))=\Lambda^{-1}(x)s^{-1} (x) A_0(x) s(x)\Lambda(x) + \Lambda^{-1}(x)s^{-1} (x) ds(x)\Lambda(x) + \Lambda^{-1}(x)d\Lambda(x)\,.
\label{ageneral}
\een
Once the zero section $s(x)$ has been fixed, there is still the freedom of $ SO(d-1,1)$ transformations, corresponding to different choices of $\Lambda(x)$ in (\ref{ageneral}). These are the standard local Lorentz transformations in the vielbein formalism of general relativity.

The choice of zero section induces a local splitting of the $\frak{so}(d,1)$ connection, according to
\ben
\frak{so}(d,1) \simeq \frak{so}(d,1)/\frak{so}(d-1,1) \oplus \frak{so}(d-1,1)\,.
\label{splitting}
\een
This splitting is invariant under the adjoint action of $ SO(d-1,1)$, thus the different parts of the connection will not mix under $ SO(d-1,1)$ transformations. The geometry is said to be {\it reductive}.

Because we assume $A$ to be a Cartan connection, the $\frak{so}(d,1)/\frak{so}(d-1,1)$ part acts as a soldering form, corresponding to the standard vielbein of general relativity; in particular, ${e_{\mu}}^i$ is an invertible matrix. The soldering form maps vectors in the tangent space $T_x \M$ at a point $x$ in spacetime to vectors in the tangent space $T_{p(x)} \dS$ at $p(x)$ in the internal de Sitter space, given in components with respect to an orthonormal basis at $p(x)$. The vielbein that maps between the components of a vector in the orthonormal basis and the coordinate-induced basis is given in (\ref{vierbein}).

\section*{Acknowledgements}

SG is supported by EPSRC and Trinity College, Cambridge. We would like to thank Derek Wise for helpful comments regarding the first preprint of this paper, and the referees for suggestions for rearrangement of an earlier version.

\end{document}